\documentclass[%
preprint,
 amsmath,amssymb,
]{revtex4-1}

\usepackage{graphicx,color}
\usepackage{dcolumn}
\usepackage{bm}

\begin{document}



%
%
%

\title{Tunable excitonic insulator in quantum limit graphite}
\author{Z.~Zhu,$^{1,2,*}$ R.~D.~McDonald,$^1$ A. Shekhter$^{1,3}$ B.~J.~Ramshaw,$^1$  K.~A.~Modic,$^1$ F.~F.~Balakirev$^1$ and N.~Harrison$^1$}

\affiliation{$^1$MS-E536, NHMFL, Los Alamos National Laboratory, Los Alamos, New Mexico 87545\\
$^2$Wuhan National High Magnetic Field Center, School of Physics, Huazhong University of Science and Technology, 1037 Luoyu Road, 430074 Wuhan, China\\
$^3$National High Magnetic Field Laboratory, Florida State University, 1800~E.~Paul Dirac Dr., Tallahassee, Florida 32310\\}

\date{\today}


\pacs{Valid PACS appear here}
\maketitle



{\bf Half a century ago, Mott noted that tuning the carrier density of a semimetal towards zero produces an insulating state in which electrons and holes form bound pairs~\cite{mott1}. It was later argued that such pairing persists even if a semiconducting gap opens in the underlying band structure, giving rise to what has become known as the strong coupling limit of an `excitonic insulator~\cite{knox1}.' While these `weak' and `strong' coupling extremes were subsequently proposed to be manifestations of the same excitonic state of electronic matter~\cite{jerome1,halperin1,comte1}, the predicted continuity of such a phase across a band gap opening has not been realized experimentally in any material. Here we show the quantum limit of graphite~\cite{mcclure1,nakao1,takada1}, by way of temperature and angle-resolved magnetoresistance measurements, to host such an excitonic insulator phase that evolves continuously between the weak and  strong coupling limits. We find that the maximum transition temperature $T_{\rm EI}$ of the excitonic phase is coincident with a band gap opening in the underlying electronic structure at $B_0=$~46~$\pm$~1~T, which is evidenced above $T_{\rm EI}$ by a thermally broadened inflection point in the magnetoresistance. 
The overall asymmetry of the observed phase boundary around $B_0$ closely matches theoretical predictions of a magnetic field-tuned excitonic insulator phase~\cite{jerome1,halperin1,comte1,abrikosov1,abrikosov2} in which the opening of a band gap marks a crossover from predominantly momentum-space pairing to real-space pairing~\cite{jerome1,halperin1,comte1}}.

\begin{figure}[ht!!!!!!!!!!] 
\includegraphics[width=15cm]{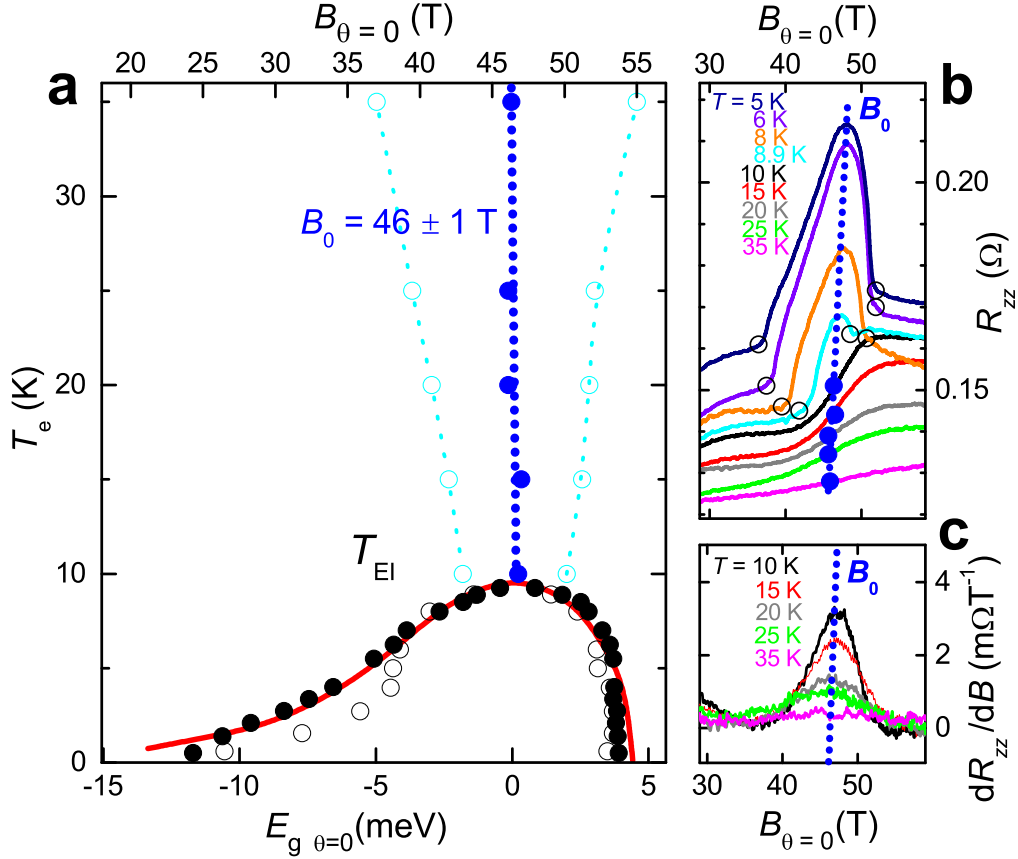}
\caption{{\bf Fixed angle magnetic phase diagram of graphite.} 
({\bf a}), Experimental phase boundary at $\theta=$~0$^\circ$ (filled black circles) extracted from $R_{xx}$ (see Supplementary Information) and the onset of insulating behavior (open black circles) extracted from $R_{zz}$, plotted against $B$ (top axis) and $E_{\rm g}$ according to Equation~\ref{gap} with $E_0=$~24.4~meV (bottom axis). The red line indicates the mean field phase boundary of an excitonic insulator phase at high magnetic fields according to Abrikosov~\cite{abrikosov1}.  ({\bf b}), $R_{zz}$ in the vicinity of the excitonic insulator phase at selected temperatures as indicated. ({\bf c}), The derivative $\partial R_{zz}/\partial B$ for $T>T_{\rm EI}$. In all panels, filled blue circles indicate the point of inflection in $R_{zz}$ at $T>T_{\rm EI}$ in graphite (where $T_{\rm EI}\approx$~9.3~K), with a blue dotted line providing a guide to the eye. The cyan circles (and dotted line) indicate the half-maximum width of $\partial R_{\rm zz}/\partial B$.
}
\label{diagram}
\end{figure}

\begin{figure}
\includegraphics[width=12cm]{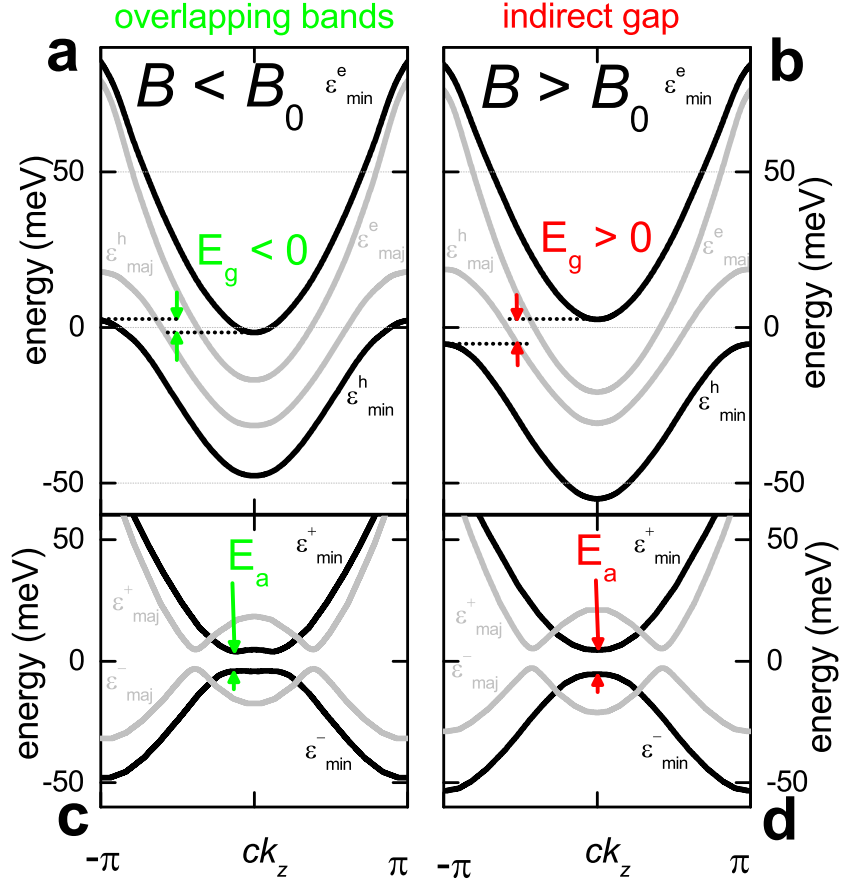}
\caption{{\bf Ultraquantum limit electronic structure of graphite in the proximity to a gap opening in the minority-spin bands}. 
({\bf a}), Electronic dispersion of the Landau subbands in graphite according to Takada and Goto~\cite{takada1} at $B<B_0$, giving rise to a small overlap ($E_{\rm g}<0$) between the minority-spin electron ($\varepsilon^{\rm e}_{\rm min}$) and hole ($\varepsilon^{\rm h}_{\rm min}$) Landau subbands (depicted in black). The majority-spin bands ($\varepsilon^{\rm e}_{\rm maj}$) and ($\varepsilon^{\rm h}_{\rm maj}$) are depicted in grey. ({\bf b}), Electronic dispersion at $B>B_0$, giving rise to a small gap ($E_{\rm g}>0$) between the minority-spin electron and hole bands. ({\bf c}), Schematic dispersion for a spin-triplet excitonic insulator phase (a spin-density-wave for weak coupling that doubles the c-axis unit cell) for $E_{\rm g}<0$. The folded dispersion is calculated from the anticrossing of the translated bands with the exciton gap function $\Delta$ using $\varepsilon^\pm_{\rm min,maj}=\frac{1}{2}[\varepsilon^{\rm e}_{\rm min,maj}(k_z)+\varepsilon^{\rm h}_{\rm min,maj}(k_z+Q_z)]\pm\sqrt{\frac{1}{4}[\varepsilon^{\rm e}_{\rm min,maj}(k_z)-\varepsilon^{\rm h}_{\rm min,maj}(k_z+Q_z)]^2+\Delta^2}$. ({\bf d}), Same as ({\bf c}) but for $E_{\rm g}>0$} 
\label{schematic}
\end{figure}

The sharp phase transitions in quantum limit graphite above 20~T~(see Fig.~\ref{diagram}) have been the subject of numerous experimental studies~\cite{iye1,iye2,yaguchi1,fauque1,kazuto1}, and have been associated with the formation of a field-induced density-wave phase~\cite{yoshioka1,sugihara1,takahashi1,takada1}. Despite decades having passed since its discovery~\cite{iye1}, the relationship of the density-wave phase to the opening of a band gap in the underlying electronic structure has remained unknown.  In the absence of a direct measurement of the underlying gap, it has been assumed from fixed angle studies performed thus far (i.e. $\theta=$~0$^\circ$)~\cite{yaguchi1,fauque1,kazuto1}  that a band gap opening coincides with the upper magnetic field phase boundary of the phase near $\approx$~54~T~\cite{takada1} (see Fig.~\ref{diagram}a). Such an analysis suggests that the entire magnetic field-induced phase lies on the weak coupling side where Landau subbands always overlap (i.e. Fig.~\ref{schematic}a)~\cite{yoshioka1} and where pairs are formed by connecting opposing momentum-states on the Fermi surface.

Rather than being coincident with the upper magnetic field boundary of the phase, we show here that the band gap opening in the underlying electronic structure coincides with the magnetic field at which the transition temperature is maximum, therefore exhibiting the signature characteristics of an excitonic insulator phase~\cite{jerome1,halperin1,comte1,abrikosov1}. 
The key experimental evidence for the gap opening at a magnetic field substantially below the upper boundary of the field-induced density-wave phase is provided by a point of inflection in the inter-plane electrical resistance $R_{zz}$ at temperatures above the ordering transition temperature $T_{\rm EI}\approx$~9.3~K (indicated by solid blue circles in Fig.~\ref{diagram}). 
The thermal evolution of the width of the peak in the derivative $\partial R_{zz}/\partial B$ in Fig.~\ref{diagram} shows that this point of inflection becomes increasingly sharp on lowering the temperature towards $T_{\rm EI}$, indicating a discontinuity at $B_0$ in the underlying band structure at low temperatures. 
Once density-wave ordering sets in at temperatures below $T_{\rm EI}$, the inflection point gives way to insulating behavior in $R_{zz}$ within the ordered phase (onset indicated by open circles in Fig.~\ref{diagram}). The complete phase boundary (solid black circles in Fig.~\ref{diagram}a) is traced from both inter-plane (see Fig.~\ref{diagram}b) and in-plane resistance (see Supplementary Information) data.

\begin{figure}
\includegraphics[width=12cm]{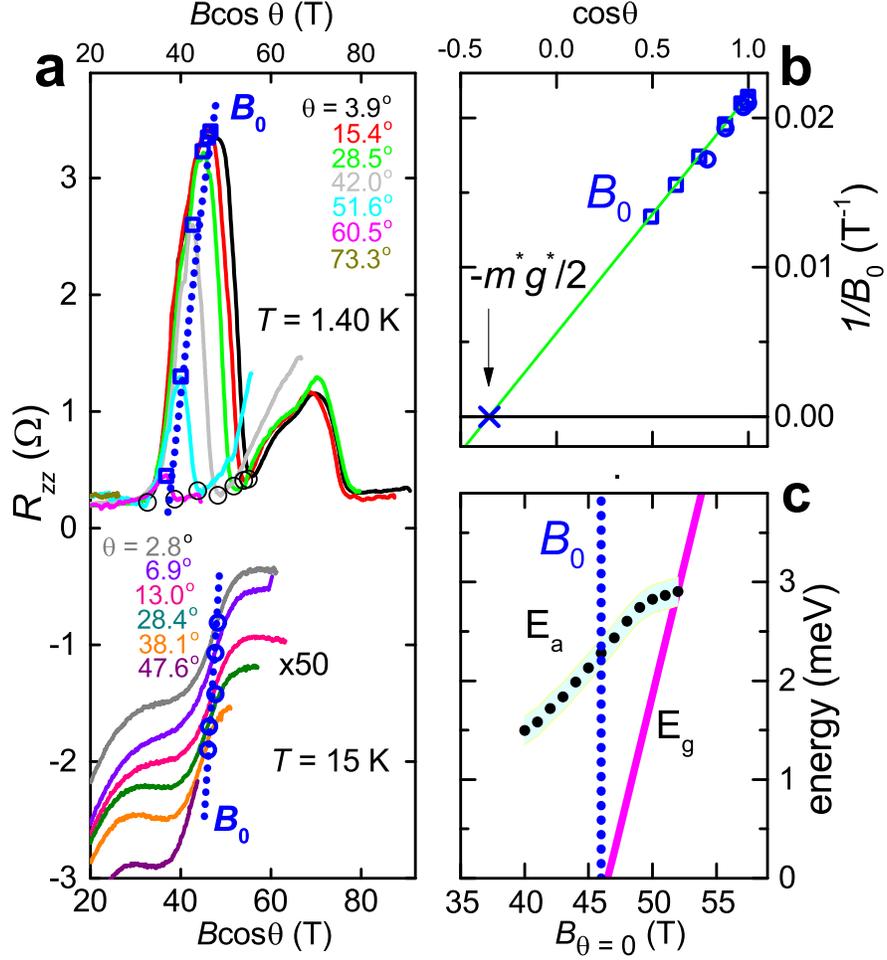}
\caption{{\bf Angle-dependent magnetoresistance data.} 
({\bf a}), $R_{zz}$ at $T=$~1.40~K and $T=$~15~K at several different angles $\theta$ as indicated (the 15~K curves have been shifted and rescaled for clarity). Open black circles indicate the onset of insulating behavior (for $T\leq T_{\rm EI}$).
({\bf b}), A plot of $1/B_0$ versus $\cos\theta$, yielding $m^\ast g^\ast$ from the intercept of a fit to Equation 1 (green line). In panels ({\bf a}) and ({\bf b}), the point of inflection for $T>T_{\rm EI}$ (blue open circles) coincides with the peak in $R_{zz}$ (blue open squares) for $T<T_{\rm EI}$. Blue dotted lines are a guide to the eye. ({\bf c}), A comparison of the field dependence of the energy gap $E_{\rm a}$ (circles with grey shading indicating the error bars) extracted from thermally activated transport measurements~\cite{fauque1} (see Supplementasry Information) with the band gap $E_{\rm g}$ (magenta). 
}
\label{data}
\end{figure}

We further use spin and orbital tuning to precisely determine the nature of the band gap opening in graphite. Because each of these contributions depends differently on the orientation of the magnetic field in layered materials, our angle-resolved measurements (see Fig.~\ref{data}) enable selective tuning of the spin and orbital contributions to the band gap. From such tuning, we infer the gap opening to occur between minority-spin electron and hole Landau subbands (shown in Fig.~\ref{schematic}b).
The inflection in $R_{zz}$ (which also coincides in field with the maximum in $R_{zz}$ within the excitonic insulator phase in Figs.~\ref{diagram}b and \ref{data}a) is observed to shift in field on rotating the polar angle $\theta$ between the magnetic field and the crystalline $c$-axis. The angle-dependence of the inflection matches the expected behavior for the opening of this band gap 
\begin{equation}\label{gap}
E_{\rm g}=\frac{\hbar e}{m^\ast}B\cos\theta+g^\ast\mu_{\rm B}B-E_0
\end{equation}
between the lowest Landau levels of minority-spin electrons and holes due to the competition between quasi-two-dimensional Landau quantization and isotropic Zeeman splitting. Here, $E_{\rm g}$ is positive for $B>B_0$ and negative (corresponding to a band overlap) for $B<B_0$ (see schematic in Figs.~\ref{schematic}a and b). The first term on the right-hand-side (in which $m^\ast$ is an effective mass for the first Landau level) results from orbital quantization within the two-dimensional honeycomb layers, the second term (in which $g^\ast$ is the effective {\it g}-factor, which is approximately isotropic in graphite, and $\mu_{\rm B}$ is the Bohr magneton) results from the Zeeman coupling of the magnetic field to the electron spin while the third ($E_0$) is a constant relating to the inter-plane electronic band structure of graphite. 

Defining $B_0$ as the field at which the band gap opens (i.e. $E_{\rm g}=0$), Equation~\ref{gap} produces a linear dependence of $1/B_0$ on $\cos\theta$, with an offset of $-(m^\ast/m_{\rm e})g^\ast/2$. On plotting $1/B_0$ data versus $\cos\theta$ in Fig.~\ref{data}b, the intercept of the fitted green line yields $(m^\ast/m_{\rm e})g^\ast/2\approx$~(0.35~$\pm$~0.01) (where $m_{\rm e}$ is the free electron mass), which is the same (to within experimental uncertainty) as the value $\approx0.37$ obtained from the known parameters of graphite ($g^\ast=$~2.5~\cite{schneider1} and $(m^\ast/m_{\rm e})=0.3$~\cite{nakao1,takada1,footnote1}).

Our measurements identify the band gap opening in the underlying electronic structure to coincide with the maximum $T_{\rm EI}$ of the asymmetric excitonic phase boundary (black circles in Fig.~\ref{diagram}a), consistent with theoretical calculations~\cite{abrikosov1} (red line). Electron-hole pairing for $B<B_0$ occurs at the Fermi surface in momentum-space in accordance with a conventional weakly coupled spin- or charge-density wave phase~\cite{yoshioka1,sugihara1,takahashi1} (schematic in Fig.~\ref{schematic}c). Such behavior has been confirmed experimentally by the observation of an exponential increase in the transition temperature as the gap increases with increasing magnetic field~\cite{iye3}.  At $B\approx B_0$, however, singularities in the electronic density-of-states at the top of the minority-spin hole band and bottom of the minority-spin electron band coincide with the chemical potential, causing strongly bound minority-spin pairs to greatly outnumber weakly bound majority-spin pairs and therefore dominate the thermodynamics. When $B>B_0$, the minority-spin pairing takes place across a band gap, thereby becoming truly excitonic in nature~\cite{knox1} (schematic in Fig.~\ref{schematic}d). Pairing across a gap is predicted to give rise to an increasingly dilute density of excitons as the magnetic field is increased~\cite{comte1,abrikosov1,halperin1,jerome1}. The exciton gap function, $\Delta$, is expected to approach zero at the upper extremity of the phase (near $\approx$~54~T in Fig.~\ref{diagram}a). The total energy gap, which will determine the thermally activated transport properties of such a correlated electron state, is given by the band gap and correlation gap added in quadrature $E_{\rm a} =\sqrt{E_{\rm g}^2+\Delta^2}$. This gap becomes comparable to the band gap $E_{\rm g}$ when the exciton density vanishes~\cite{comte1,halperin1,abrikosov1,jerome1}. Such behavior is demonstrated in Fig.~\ref{data}c by the evolution of an activation gap within the excitonic insulator phase, obtained from Arrhenius plots of $R_{zz}$~\cite{fauque1} (see Supplementary Information), that continues to grow in the region $B>B_0$, and then intersects with $E_{\rm g}$ on approaching the upper phase boundary. The point of intersection provides a lower bound estimate of ~$\approx$~3~meV for the exciton binding potential energy, which is expected to be similar to the value of $\Delta$ at the peak transition temperature~\cite{halperin1,jerome1,comte1}.

\begin{figure}
\includegraphics[width=10cm]{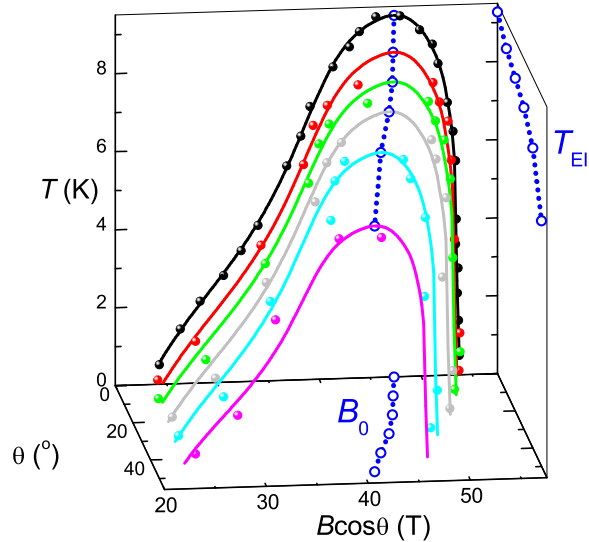}
\caption{{\bf Angle-dependent phase diagram of graphite.} 
Evolution of the excitonic phase boundary with field rotation angle $\theta$, where solid lines represent a spine fit to the phase boundary at $\theta=0$ which for $\theta>0$ has been rescaled as a guide as a guide to the eye. Blue circles connected by dotted lines represent the interpolated optimal transition temperature at each field angle, which has further been projected onto the $T$ -- $\theta$ and $B\cos\theta$ -- $\theta$ planes, labelled as $T_{\rm EI}$ and $B_0$, respectively. The optimal transition temperature follows approximately linear dependence with $B\cos\theta$, (not immediately apparent when plotted versus $\theta$). Accompanying this weakening of the optimal $T_{\rm EI}$ is shrinking of the region of the excitonic phase around $B_0\cos\theta$ as we rotate field into the honeycomb plane. The transition temperature is expected to collapse to zero at all fields for angles above 70$^\circ$ (corresponding to a magnetic field of  $\sim$~100~T). 
}
\label{interactions}
\end{figure}

The stability of the excitonic insulator phase centered around $B_0$ depends on the effective strength of the interactions determining the binding energy. 
The combination of anisotropic orbital and isotropic Zeeman contributions to $E_{\rm g}$ (as defined by Equation~\ref{gap}) shifts the the opening of the band gap and hence the optimal transition temperature of the exitonic insulator to lower values of the component of magnetic field perpendicular to the planes, $B_0\cos\theta$, as $\theta$ is increased (Figs.~\ref{data} and \ref{interactions}). 
The reduced optimal transition temperature of the excitonic insulator phase and its reduced extent in $B\cos\theta$ at higher angles suggest that the pairing strength is inversely correlated with the magnetic length at a given particle density (the magnetic length being inversely proportional to $B\cos\theta$ and the particle density depending only on $E_{\rm g}$). The angle-dependent measurements hence provide an experimental means of tuning the pairing  strength in a condensed matter system, independent of the electron gas density, analogous to that achieved in cold atomic gases~\cite{bloch1}. We speculate that the observed dependence of the effective binding potential on in-plane magnetic length is a result of the poor screening of the Coulomb interactions at low density, especially in the vicinity of $E_g=0$, in line with the original discussion of an excitonic transition by Mott~\cite{mott1}.

While the nature of the broken symmetry in quantum-limit graphite has remained an open question ~\cite{sugihara1,yaguchi1,fauque1,takahashi1,yoshioka1,kazuto1}, our observation of the maximum transition temperature at the field $B_0$ implies that the broken symmetry accompanying its formation bridges opposing limits of the phase diagram in which excitons are strongly and weakly bound. 
Beyond the proposed formation of a density-wave in the low-field, weak-coupling limit, the possibilities for the broken symmetry in the excitonic phase include a Bose-Einstein condensate of excitons~\cite{comte1}, a Wigner crystalline~\cite{halperin1} or supersolid~\cite{joglekar1} state of excitons, or a state with chiral symmetry breaking~\cite{khveshchenko1}. One way of forming a reconstructed electronic dispersion~\cite{takada1} typical of an excitonic phase~\cite{jerome1,abrikosov1,halperin1} is a spin-ordered phase with an inter-plane component to the ordering vector of $Q_z=\pi/c$ (shown schematically in Fig.~\ref{schematic}c and d to couple electrons and holes of opposite spin). 
This has the attractive property of producing broken translational symmetry along the $c$-axis, as expected for a crystalline exciton phase~\cite{halperin1}, while leaving the in-plane mobility of the electrons and holes intact and open to the possibility of superfluid~\cite{comte1} or supersolid behavior~\cite{joglekar1}. This same $Q_z$ vector also nests the majority-spin bands, which is likely to be important for driving the fully-insulating state. 

Our observation of an ordered excitonic phase nucleating around the opening of a band gap, suggests that graphite is an attractive material for investigating exotic ordered states in ultra-low density electronic systems~\cite{mcclure1,takada1,kotov1} with poorly screened coulomb interactions~\cite{mott1}. The nature of the broken symmetry in the excitonic insulator phase and whether the onset of the insulating phase precedes or is coincident with it remains an open question. 
In particular there exists a second field-induced phase at higher magnetic fields, as reported by Fauqu\'{e} {\it et al}~\cite{fauque1}, raising the possibility that this is a second excitonic insulator phase involving only the majority-spin carriers (the upper phase between $\approx$~55 and 75~T also being evident in Fig.~\ref{data}c). The similarity in shape of the second magnetic field-induced phase to that at low fields suggest that it may be centered around a band gap opening between the majority-spin Landau subbands at $\approx$~70~T.

Z.Z. and R.D.M. thank K.~Behnia and B.~Fauqu\'{e} for fruitful discussions while N.H. and R.D.M.  thank A.~V.~Balatsky for useful discussions. This research performed under the DOE BES `Science at 100~tesla' at the magnet lab. which is supported by NSF Cooperative Agreement No. DMR-1157490.


\begin{thebibliography}{10}
\bibitem{footnote1} This effective mass parameter corresponds to the magnetic field-dependence of the energy difference between $n=0$ (electron) and $n=-1$ (hole) Landau levels~\cite{takada1}, and is larger than the effective mass ($\approx$~0.05~$m_{\rm e}$) of the electron and hole pockets.
\end{thebibliography}
\end{document}